%% file: main.tex
\pdfoutput=1
\documentclass[12pt,a4paper]{article}
\usepackage{ifthen} % for conditional statements
\newboolean{pdflatex}
\setboolean{pdflatex}{true} % False for eps figures 

\newboolean{articletitles}
\setboolean{articletitles}{true} % False removes titles in references

\newboolean{uprightparticles}
\setboolean{uprightparticles}{false} %True for upright particle symbols

\def\paperauthors{LHCb collaboration} % Leave as is for PAPER, CONF and FIGURE
\def\paperasciititle{Observation of new excited Bs states} % Set ASCII title here !! MAKE sure it's only ASCII characters !! 
\def\papertitle{Observation of new excited \Bs states} % Latex formatted title
\def\paperkeywords{{High Energy Physics}, {LHCb}} % Comma separated list
\def\papercopyright{\the\year\ CERN for the benefit of the LHCb collaboration} % new since 9/Apr/2018
\def\paperlicence{CC-BY-4.0 licence}
\def\paperlicenceurl{https://creativecommons.org/licenses/by/4.0/}

\input{preamble}
\usepackage{subcaption}
\usepackage[separate-uncertainty=true]{siunitx}
\DeclareSIUnit\evolt{e\kern -0.1em V}
\usepackage{physics}
\usepackage{booktabs}
\usepackage[capitalize]{cleveref}
\usepackage{multirow}

\input{macros}

\usepackage{longtable} % only for template; not usually to be used in PAPERs

\begin{document}

%%%%%%%%%%%%%%%%%%%%%%%%%
%%%%% Title     %%%%%%%%%
%%%%%%%%%%%%%%%%%%%%%%%%%
\renewcommand{\thefootnote}{\fnsymbol{footnote}}
\setcounter{footnote}{1}

% %%%%%%% CHOOSE TITLE PAGE--------
%\onecolumn
%\input{title-LHCb-INT}
%\input{title-LHCb-ANA}
%\input{title-LHCb-CONF}
%\input{title-LHCb-FIGURE}
\input{title-LHCb-PAPER}

%\twocolumn
% %%%%%%%%%%%%% ---------

\renewcommand{\thefootnote}{\arabic{footnote}}
\setcounter{footnote}{0}

%%%%%%%%%%%%%%%%%%%%%%%%%%%%%%%%
%%%%%  Table of Content   %%%%%%
%%%%%%%%%%%%%%%%%%%%%%%%%%%%%%%%
%%%% Uncomment if desired
%\tableofcontents
\cleardoublepage

%%%%%%%%%%%%%%%%%%%%%%%%%
%%%%% Main text %%%%%%%%%
%%%%%%%%%%%%%%%%%%%%%%%%%

\pagestyle{plain} % restore page numbers for the main text
\setcounter{page}{1}
\pagenumbering{arabic}

%% Uncomment during review phase. 
%% Comment before a final submission.
%% \linenumbers

\input{Introduction}

\input{Selection}
\input{FitDescription}

\input{Production}
\input{Conclusion}
\input{acknowledgements}

\addcontentsline{toc}{section}{References}
%\setboolean{inbibliography}{true}
\bibliographystyle{LHCb}
\bibliography{main}%,standard,LHCb-PAPER,LHCb-CONF,LHCb-DP,LHCb-TDR}
 
 %% \newpage
 %% \input{Supplementary}
 
\newpage
\input{LHCb_Authorship_23-Jul-2020}

\end{document}

%% file: preamble.tex
\usepackage[top=1in, bottom=1.25in, left=1in, right=1in]{geometry}

\usepackage{microtype}
\usepackage{lineno}  % for line numbering during review
\usepackage{xspace} % To avoid problems with missing or double spaces after
\usepackage{caption} %these three command get the figure and table captions automatically small

\usepackage{graphicx}  % to include figures (can also use other packages)
\usepackage{color}
\usepackage{colortbl}
\graphicspath{{./figs/}} % Make Latex search fig subdir for figures
\usepackage{amsmath} % Adds a large collection of math symbols
\usepackage{amssymb}
\usepackage{amsfonts}
\usepackage{upgreek} % Adds in support for greek letters in roman typeset

\newcommand*\patchAmsMathEnvironmentForLineno[1]{%
\expandafter\let\csname old#1\expandafter\endcsname\csname #1\endcsname
\expandafter\let\csname oldend#1\expandafter\endcsname\csname
end#1\endcsname
 \renewenvironment{#1}%
   {\linenomath\csname old#1\endcsname}%
   {\csname oldend#1\endcsname\endlinenomath}%
}
\newcommand*\patchBothAmsMathEnvironmentsForLineno[1]{%
  \patchAmsMathEnvironmentForLineno{#1}%
  \patchAmsMathEnvironmentForLineno{#1*}%
}
\AtBeginDocument{%
\patchBothAmsMathEnvironmentsForLineno{equation}%
\patchBothAmsMathEnvironmentsForLineno{align}%
\patchBothAmsMathEnvironmentsForLineno{flalign}%
\patchBothAmsMathEnvironmentsForLineno{alignat}%
\patchBothAmsMathEnvironmentsForLineno{gather}%
\patchBothAmsMathEnvironmentsForLineno{multline}%
\patchBothAmsMathEnvironmentsForLineno{eqnarray}%
}

\usepackage{hyperxmp}

\usepackage[pdftex,
            pdfauthor={\paperauthors},
            pdftitle={\paperasciititle},
            pdfkeywords={\paperkeywords},
            pdfcopyright={Copyright (C) \papercopyright},
            pdflicenseurl={\paperlicenceurl}]{hyperref}
\usepackage[colorinlistoftodos,textsize=scriptsize]{todonotes}

\usepackage[bottom,flushmargin,hang,multiple]{footmisc}

\usepackage[all]{hypcap} % Internal hyperlinks to floats.

\input{lhcb-symbols-def} % Add in the predefined LHCb symbols

\usepackage{cite} % Allows for ranges in citations
\usepackage{mciteplus}

%% file: lhcb-symbols-def.tex
\usepackage{xspace} 
\usepackage{upgreek}

\def\lhcb   {\mbox{LHCb}\xspace}

\def\MagUp {\mbox{\em Mag\kern -0.05em Up}\xspace}

\ifthenelse{\boolean{uprightparticles}}%
{

 \def\Pmu         {\ensuremath{\upmu}\xspace}

 \def\Ppi         {\ensuremath{\uppi}\xspace}

 \def\Ppsi        {\ensuremath{\uppsi}\xspace}

 \def\PDelta      {\ensuremath{\Delta}\xspace}                 
 \def\PXi         {\ensuremath{\Xi}\xspace}                 
 \def\PLambda     {\ensuremath{\Lambda}\xspace}                 
 \def\PSigma      {\ensuremath{\Sigma}\xspace}                 
 \def\POmega      {\ensuremath{\Omega}\xspace}                 
 \def\PUpsilon    {\ensuremath{\Upsilon}\xspace}

 \def\PB      {\ensuremath{\mathrm{B}}\xspace}                 
                  
 \def\PD      {\ensuremath{\mathrm{D}}\xspace}

 \def\PJ      {\ensuremath{\mathrm{J}}\xspace}                 
 \def\PK      {\ensuremath{\mathrm{K}}\xspace}

 \def\Pb      {\ensuremath{\mathrm{b}}\xspace}                 
 \def\Pc      {\ensuremath{\mathrm{c}}\xspace}

 \def\Pi      {\ensuremath{\mathrm{i}}\xspace}

 \def\Ps      {\ensuremath{\mathrm{s}}\xspace}

 \def\thebaroffset{0.0em}
}
{

 \def\Pmu         {\ensuremath{\mu}\xspace}

 \def\Ppi         {\ensuremath{\pi}\xspace}

 \def\Ppsi        {\ensuremath{\psi}\xspace}                 
                  
 \mathchardef\PDelta="7101
 \mathchardef\PXi="7104
 \mathchardef\PLambda="7103
 \mathchardef\PSigma="7106
 \mathchardef\POmega="710A
 \mathchardef\PUpsilon="7107
                  
 \def\PB      {\ensuremath{B}\xspace}                 
                  
 \def\PD      {\ensuremath{D}\xspace}

 \def\PJ      {\ensuremath{J}\xspace}                 
 \def\PK      {\ensuremath{K}\xspace}

 \def\Pb      {\ensuremath{b}\xspace}                 
 \def\Pc      {\ensuremath{c}\xspace}

 \def\Pi      {\ensuremath{i}\xspace}

 \def\Ps      {\ensuremath{s}\xspace}

 \def\thebaroffset{0.18em}
}
\newcommand{\offsetoverline}[2][\thebaroffset]{\kern #1\overline{\kern -#1 #2}}%

\makeatletter
\ifcase \@ptsize \relax% 10pt
  \newcommand{\miniscule}{\@setfontsize\miniscule{4}{5}}% \tiny: 5/6
\or% 11pt
  \newcommand{\miniscule}{\@setfontsize\miniscule{5}{6}}% \tiny: 6/7
\or% 12pt
  \newcommand{\miniscule}{\@setfontsize\miniscule{5}{6}}% \tiny: 6/7
\fi
\makeatother

\DeclareRobustCommand{\optbar}[1]{\shortstack{{\miniscule (\rule[.5ex]{1.25em}{.18mm})}
  \\ [-.7ex] $#1$}}

\def\mup        {{\ensuremath{\Pmu^+}}\xspace}
\def\mun        {{\ensuremath{\Pmu^-}}\xspace} % muon negative (\mum is taken)

\def\squark    {{\ensuremath{\Ps}}\xspace}

\def\cquark    {{\ensuremath{\Pc}}\xspace}

\def\bquark    {{\ensuremath{\Pb}}\xspace}

\def\pion   {{\ensuremath{\Ppi}}\xspace}

\def\pip    {{\ensuremath{\pion^+}}\xspace}
\def\pim    {{\ensuremath{\pion^-}}\xspace}

\def\kaon    {{\ensuremath{\PK}}\xspace}
\def\KorKbar {\kern \thebaroffset\optbar{\kern -\thebaroffset \PK}{}\xspace}

\def\Kp      {{\ensuremath{\kaon^+}}\xspace}
\def\Km      {{\ensuremath{\kaon^-}}\xspace}

\def\Dbar    {{\ensuremath{\offsetoverline{\PD}}}\xspace}
\def\D       {{\ensuremath{\PD}}\xspace}

\def\DorDbar {\kern \thebaroffset\optbar{\kern -\thebaroffset \PD}\xspace}

\def\Dzb     {{\ensuremath{\Dbar{}^0}}\xspace}
\def\Dp      {{\ensuremath{\D^+}}\xspace}
\def\Dm      {{\ensuremath{\D^-}}\xspace}

\def\DpDm    {\ensuremath{\Dp {\kern -0.16em \Dm}}\xspace}

\def\B       {{\ensuremath{\PB}}\xspace}

\def\BorBbar {\kern \thebaroffset\optbar{\kern -\thebaroffset \PB}\xspace}

\def\Bd      {{\ensuremath{\B^0}}\xspace}

\def\BdorBdbar {\kern \thebaroffset\optbar{\kern -\thebaroffset \Bd}\xspace}
\def\Bu      {{\ensuremath{\B^+}}\xspace}

\def\Bp      {{\ensuremath{\Bu}}\xspace}

\def\Bs      {{\ensuremath{\B^0_\squark}}\xspace}

\def\BsorBsbar {\kern \thebaroffset\optbar{\kern -\thebaroffset \Bs}\xspace}

\def\jpsi     {{\ensuremath{{\PJ\mskip -3mu/\mskip -2mu\Ppsi}}}\xspace}

\def\Y#1S{\ensuremath{\PUpsilon{(#1S)}}\xspace}

\def\LorLbar     {\kern \thebaroffset\optbar{\kern -\thebaroffset \PLambda}\xspace}

\def\BF         {{\ensuremath{\mathcal{B}}}\xspace}

\newcommand{\decay}[2]{\ensuremath{#1\!\to #2}\xspace} 

\def\to                 {\ensuremath{\rightarrow}\xspace}

\def\AT#1     {\ensuremath{A_{\mathrm{T}}^{#1}}\xspace}           % 2

\def\C#1      {\ensuremath{\mathcal{C}_{#1}}\xspace}                       % 9
\def\Cp#1     {\ensuremath{\mathcal{C}_{#1}^{'}}\xspace}                    % 7
\def\Ceff#1   {\ensuremath{\mathcal{C}_{#1}^{\mathrm{(eff)}}}\xspace}        % 9  
\def\Cpeff#1  {\ensuremath{\mathcal{C}_{#1}^{'\mathrm{(eff)}}}\xspace}       % 7
\def\Ope#1    {\ensuremath{\mathcal{O}_{#1}}\xspace}                       % 2
\def\Opep#1   {\ensuremath{\mathcal{O}_{#1}^{'}}\xspace}                    % 7

\newcommand{\nospaceunit}[1]{\ensuremath{\text{#1}}}       
\newcommand{\aunit}[1]{\ensuremath{\text{\,#1}}}       
\newcommand{\tev}{\aunit{Te\kern -0.1em V}\xspace}
\newcommand{\gev}{\aunit{Ge\kern -0.1em V}\xspace}
\newcommand{\mev}{\aunit{Me\kern -0.1em V}\xspace}
\newcommand{\kev}{\aunit{ke\kern -0.1em V}\xspace}
\newcommand{\ev}{\aunit{e\kern -0.1em V}\xspace}
 
\newcommand{\mevc}{\ensuremath{\aunit{Me\kern -0.1em V\!/}c}\xspace}
\newcommand{\gevc}{\ensuremath{\aunit{Ge\kern -0.1em V\!/}c}\xspace}
\newcommand{\mevcc}{\ensuremath{\aunit{Me\kern -0.1em V\!/}c^2}\xspace}
\newcommand{\gevcc}{\ensuremath{\aunit{Ge\kern -0.1em V\!/}c^2}\xspace}
\def\mum  {\ensuremath{\,\upmu\nospaceunit{m}}\xspace}

\def\fb   {\ensuremath{\aunit{fb}}\xspace}
\def\invfb   {\ensuremath{\fb^{-1}}\xspace}

\def\gsim{{~\raise.15em\hbox{$>$}\kern-.85em
          \lower.35em\hbox{$\sim$}~}\xspace}
\def\lsim{{~\raise.15em\hbox{$<$}\kern-.85em
          \lower.35em\hbox{$\sim$}~}\xspace}

\def\pt         {\ensuremath{p_{\mathrm{T}}}\xspace}

\def\ptot       {\ensuremath{p}\xspace}

\def\evtgen     {\mbox{\textsc{EvtGen}}\xspace}

\def\geant      {\mbox{\textsc{Geant4}}\xspace}

\def\pythia     {\mbox{\textsc{Pythia}}\xspace}

\def\tell1  {TELL1\xspace}
\def\ukl1   {UKL1\xspace}

%% file: macros.tex
\def\bstp{{\ensuremath{B^{*+}}}\xspace}
\def\bsst{{\ensuremath{B_s^{*0}}}\xspace}

\def\bsstst{{\ensuremath{B_s^{**0}}}\xspace}

\def\bstwost{{\ensuremath{B_{s2}^{*0}}}\xspace}

\def\bsone{{\ensuremath{B_{s1}^0}}\xspace}

%% file: title-LHCb-PAPER.tex
% ===============================================================================
% Purpose: LHCb-PAPER journal paper title page template
% Author: 
% Created on: 2010-09-25
% ===============================================================================

%%%%%%%%%%%%%%%%%%%%%%%%%
%%%%%  TITLE PAGE  %%%%%%
%%%%%%%%%%%%%%%%%%%%%%%%%
\begin{titlepage}
\pagenumbering{roman}

% Header ---------------------------------------------------
\vspace*{-1.5cm}
\centerline{\large EUROPEAN ORGANIZATION FOR NUCLEAR RESEARCH (CERN)}
\vspace*{1.5cm}
\noindent
\begin{tabular*}{\linewidth}{lc@{\extracolsep{\fill}}r@{\extracolsep{0pt}}}
\ifthenelse{\boolean{pdflatex}}% Logo format choice
{\vspace*{-1.5cm}\mbox{\!\!\!\includegraphics[width=.14\textwidth]{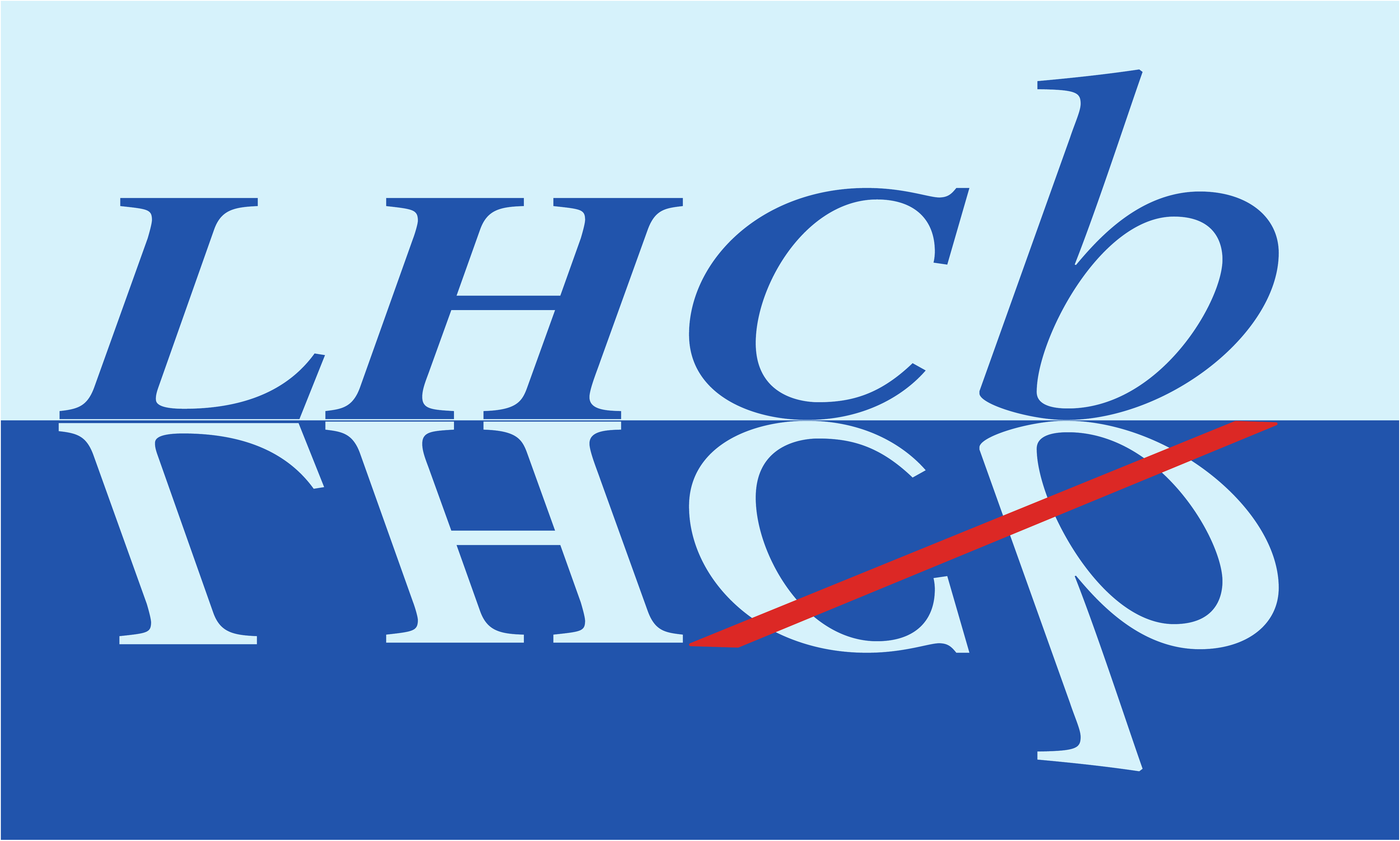}} & &}%
{\vspace*{-1.2cm}\mbox{\!\!\!\includegraphics[width=.12\textwidth]{figs/lhcb-logo.eps}} & &}%
\\
 & & CERN-EP-2020-185 \\  % ID 
 & & LHCb-PAPER-2020-026 \\  % ID 
 & & July 28, 2021 \\ % Date - Can also hardwire e.g.: 23 March 2010
% not in paper \hline
\end{tabular*}

\vspace*{4.0cm}

% Title --------------------------------------------------
{\normalfont\bfseries\boldmath\huge
\begin{center}
% DO NOT EDIT HERE. Instead edit macro in main.tex to keep metadata correct
  \papertitle 
\end{center}
}

\vspace*{1.0cm}

% Authors -------------------------------------------------
\begin{center}
%In the footnote, replace 'paper' by 'Letter' in case of submission to PRL or PLB 
% Edit macro in main.tex to keep metadata correct
\paperauthors\footnote{Authors are listed at the end of this paper.}
\end{center}

\vspace{\fill}

% Abstract -----------------------------------------------
\begin{abstract}
  \noindent
  A structure is observed in the $\B^{\pm}\kaon^{\mp}$ mass spectrum in a sample of proton--proton collisions at centre-of-mass energies of 7, 8, and \SI{13}{\tera\evolt}, collected with the LHCb detector and corresponding to a total integrated luminosity of 9\invfb. The structure is interpreted as the result of overlapping excited \Bs states. With high significance, a two-peak hypothesis provides a better description of the data than a single resonance. Under this hypothesis the masses and widths of the two states, assuming they decay directly to $\B^{\pm}\kaon^{\mp}$, are determined to be
  \begin{align*}
    m_1      &= 6063.5 \pm   1.2 \text{ (stat)} \pm   0.8\text{ (syst)}\mev, \\
\Gamma_1 &=  26 \pm   4 \text{ (stat)} \pm   4\text{ (syst)}\mev, \\
m_2      &= 6114 \pm   3 \text{ (stat)} \pm   5\text{ (syst)}\mev, \\
\Gamma_2 &=  66 \pm   18 \text{ (stat)} \pm   21\text{ (syst)}\mev.
  \end{align*}
  Alternative values assuming a decay through $\B^{*\pm}\kaon^{\mp}$, with a missing photon from the $B^{*\pm} \to B^{\pm}\gamma$ decay, which are shifted by approximately 45\mev, are also determined. The possibility of a single state decaying in both channels is also considered. The ratio of the total production cross-section times branching fraction of the new states relative to the previously observed \bstwost state is determined to be $0.87 \pm 0.15 \text{ (stat)} \pm 0.19 \text{ (syst)}$.
\end{abstract}

\vspace*{1.0cm}

\begin{center}
  Published in Eur. Phys. J. C81 
(2021) 601
\end{center}

\vspace{\fill}

{\footnotesize 
% Edit macro in main.tex to keep metadata correct
\centerline{\copyright~\papercopyright. \href{\paperlicenceurl}{\paperlicence}.}}
\vspace*{2mm}

\end{titlepage}

%%%%%%%%%%%%%%%%%%%%%%%%%%%%%%%%
%%%%%  EOD OF TITLE PAGE  %%%%%%
%%%%%%%%%%%%%%%%%%%%%%%%%%%%%%%%

%  empty page follows the title page ----
\newpage
\setcounter{page}{2}
\mbox{~}

%% file: Introduction.tex
\section{Introduction}
\label{sec:intro}

The constituent quark model~\cite{GellMann:1964nj,Zweig:352337,*Zweig:570209} has been very successful describing the quark composition and allowed quantum numbers of observed hadron states. Potential models exploiting heavy-quark symmetry~\cite{GodfreyIsgur,NEUBERT1994259,Manohar:2000dt} are used to calculate properties of mesons containing one heavy and one light quark, including those with $\bar{b}s$ quark content, termed collectively the \bsstst mesons. Yet it is still difficult to predict precise masses and widths of such states.  Experimental measurements are therefore essential. This paper presents an observation of an excess in the $\Bu\Km$ invariant-mass spectrum,\footnote{The inclusion of charge-conjugate processes is implied throughout.} which we interpret as originating from \bsstst mesons.
Two states, \bsone and \bstwost, with masses higher than the ground state \Bs and \bsst mesons have been previously observed~\cite{Aaltonen:2007ah,Abazov:2007af,LHCb-PAPER-2012-030,Aaltonen:2013atp,Sirunyan:2018grk}. Based on their masses and narrow widths, they are interpreted as two of the predicted states with orbital angular momentum $L=1$. We report the ratio of the production cross-section times branching fraction for the newly observed states relative to that of the \bstwost meson.

Predictions using a variety of methods have been made for masses and widths of states belonging to orbital or radial excitations of $\bar{b}s$ states~\cite{Godfrey:2016nwn,Ferretti:2015rsa,Lu:2016bbk,Xiao:2014ura,Sun:2014wea,Asghar:2018tha,Liu:2015lka}. The spectroscopic notation, $n^{2S+1}L_J$, is commonly used to refer to these states, where $n$ gives the radial quantum number, $S$ the sum of the quark spins, $L$ the orbital angular momentum of the quarks, and $J$ the total angular momentum.
The masses of the first radially excited states, $2^1S_0$ and $2^3S_1$, are predicted in the range between 5920\mev and 6020\mev.\footnote{Natural units with $c = 1$ are used throughout.}  Different calculations give significantly different predicted widths for these states.  In some cases they are close to 100\mev or more~\cite{Godfrey:2016nwn,Ferretti:2015rsa,Lu:2016bbk} and in others can be less than 50\mev~\cite{Xiao:2014ura,Sun:2014wea,Asghar:2018tha}. Broad states are difficult to identify experimentally, however, because of large non-resonant continuum contributions.
The first $D$-wave states ($L=2$) are predicted to have masses ranging roughly from 6050\mev to 6200\mev. The $1^3D_3$ state is almost always predicted to be relatively narrow, with a width less than 50\mev. The $J=2$ states $1^1D_2$ and $1^3D_2$ can mix, and depending on the mixing angle they may produce one of the states as narrow as 20\mev. With multiple states potentially decaying to both the $\Bu\Km$ final state directly and through a \bstp intermediate state, with a missing photon from the $B^{*+} \to B^+ \gamma$ decay, the observation of overlapping peaks in this mass region seems likely.

%% file: Selection.tex
\section{Detector and selection}

The \lhcb detector~\cite{LHCb-DP-2008-001,LHCb-DP-2014-002} is a single-arm forward
spectrometer covering the \mbox{pseudorapidity} range $2<\eta <5$,
designed for the study of particles containing \bquark or \cquark
quarks. The detector includes a high-precision tracking system
consisting of a silicon-strip vertex detector surrounding the $pp$
interaction region, a large-area silicon-strip detector located
upstream of a dipole magnet with a bending power of about
$4{\mathrm{\,Tm}}$, and three stations of silicon-strip detectors and straw
drift tubes placed downstream of the magnet.
The tracking system provides a measurement of momentum, \ptot, of charged particles with
a relative uncertainty that varies from 0.5\% at low momentum to 1.0\% at 200\gev.
The minimum distance of a track to a primary $pp$ interaction vertex (PV), the impact parameter (IP), 
is measured with a resolution of $(15+29/\pt)\mum$,
where \pt is the component of the momentum transverse to the beam, in\,\gev.
Different types of charged hadrons are distinguished using information
from two ring-imaging Cherenkov detectors. 
Photons, electrons and hadrons are identified by a calorimeter system consisting of
scintillating-pad and preshower detectors, an electromagnetic
calorimeter and a hadronic calorimeter. Muons are identified by a
system composed of alternating layers of iron and multiwire
proportional chambers.
The online event selection is performed by a trigger, 
which consists of a hardware stage, based on information from the calorimeter and muon
systems, followed by a software stage, which applies a full event
reconstruction.
At the hardware trigger stage, events are required to have a muon with high \pt or a
  hadron with high transverse energy in the calorimeter. 
  The software trigger requires a two-, three- or four-track
 vertex with a significant displacement from any PV. At least one charged particle
  must have a $\pt > 1.6\gev$ and be
  inconsistent with originating from a PV. Consistency with a PV is defined based on the $\chi^2$ difference between vertex fits including and excluding the particle under consideration.

We use data samples collected from 2011 to 2018, at centre-of-mass energies of 7, 8, and \SI{13}{\tera\evolt}, corresponding to an integrated luminosity of 9\invfb. We simulate both \bstwost decays and decays of a \bsstst meson with a mass of 300\mev above the \Bu\Km mass threshold. In the simulation, $pp$ collisions are generated using
\pythia~\cite{Sjostrand:2006za,*Sjostrand:2007gs} with a specific \lhcb
configuration~\cite{LHCb-PROC-2010-056}.  Decays of hadronic particles
are described by \evtgen~\cite{Lange:2001uf}. The
interaction of the generated particles with the detector and its response
are implemented using the \geant
toolkit~\cite{Allison:2006ve, *Agostinelli:2002hh} as described in
Ref.~\cite{LHCb-PROC-2011-006}.

The selection of the \Bu\Km candidates is performed in two steps. First, we select \Bu candidates using the decays \decay{\Bu}{\jpsi\Kp} and \decay{\Bu}{\Dzb\pip}, with \decay{\jpsi}{\mup\mun} and \decay{\Dzb}{\Kp\pim}. Each final-state particle is formed from a high-quality track that is inconsistent with being produced at any PV in the event. Each kaon, pion and muon is also required to have particle identification information consistent with its hypothesis. We do not require the candidate to have fired the trigger in the event. The selection of \Bu candidates has high purity, with a background contribution in the \Bu mass region of less than 10\%.

Subsequently, we form the \bsstst candidates by combining the \Bu candidate with a $K^-$ candidate consistent with being produced at the PV, referred to as the prompt kaon. In the same way we also select a separate sample of same-sign \Bu\Kp candidates further used for constraining combinatorial background. Strong particle-identification requirements are applied to the prompt kaons to reduce the large pion background. As observed for the known \bstwost resonance, the strongest discriminant between resonant signals and combinatorial background is the kaon transverse momentum. We therefore study the \bsstst candidates in bins of the prompt kaon \pt: $0.5 < \pt < \SI{1}{\giga\evolt}$, $1 < \pt < \SI{2}{\giga\evolt}$, and $\pt > \SI{2}{\giga\evolt}$. The \bsstst candidate mass, $m_{\Bu\Km}$, is reconstructed constraining the masses \jpsi or \Dzb meson and the \Bu meson to their known values~\cite{PDG2020} and requiring the \mbox{\Bu candidate} to have been produced at the PV.
We search for new states in a spectrum of the mass difference, $\Delta m = m_{\Bu\Km} - m_{\Bu} - m_{\Km}$, where the latter two masses refer to the known \Bp and \Km masses~\cite{PDG2020}. The $\Delta m$ spectrum, measured in the three kaon \pt regions, is shown in \cref{fig:fit}. A clear deviation from a smooth continuum shape is observed at approximately 300\mev above the mass threshold, and is consistently present in both \Bu selections.

%% file: FitDescription.tex
\section{Fit description and results}

We determine the significance of the excess using an unbinned maximum-likelihood fit to the \Bu\Km mass difference spectrum in the region from 150\mev to \SI{800}{\mega\evolt} above the kinematic threshold. The fit is performed simultaneously in three bins of the prompt kaon \pt. 

We form the background distribution from two components. The first is the combinatorial background, whose shape and yield are fixed based on a fit to the same-sign \Bu\Kp mass difference distribution using a threshold function of the form $f(\Delta m) = \Delta m^{a}\exp(-b\Delta m)$, with $a$ and $b$ as free parameters. The additional continuum background component that is not present in the same-sign sample, and which arises from associated production (AP) of the non-resonant \Bu\Km pairs, is free to vary in the fit; our model for this component is a second-degree polynomial. Given the similarity of the combinatorial and AP background shapes, statistical uncertainties in the former are expressed through the latter fit component.

The \bsstst signal is described by one or two relativistic Breit--Wigner distributions, convolved with resolution functions determined using simulation. For new states, the decay products' angular momentum ($\ell$) is unknown, making it difficult  to assign an appropriate Blatt--Weisskopf barrier factor~\cite{Blatt:1952ije}. The \bsstst state most consistently predicted to be narrow decays with a $B^{(*)+}\Km$ angular momentum of $\ell=3$, so we use this as default, but consider other angular momenta in alternative fits. The barrier radius parameter is fixed to \SI{4}{\per\giga\evolt}~\cite{LHCb-PAPER-2014-036,LHCb-PAPER-2014-067}.  When fitting with two peaks, no interference between the two new states is considered. Interference is possible if the two states are both $J=2$ mixed $D$-wave states, but in that case one of the states should be much wider, so this scenario is unlikely to produce two narrow peaks.

\begin{figure}[tbp]

	\begin{subfigure}{0.5\textwidth}
		\includegraphics[width=\textwidth]{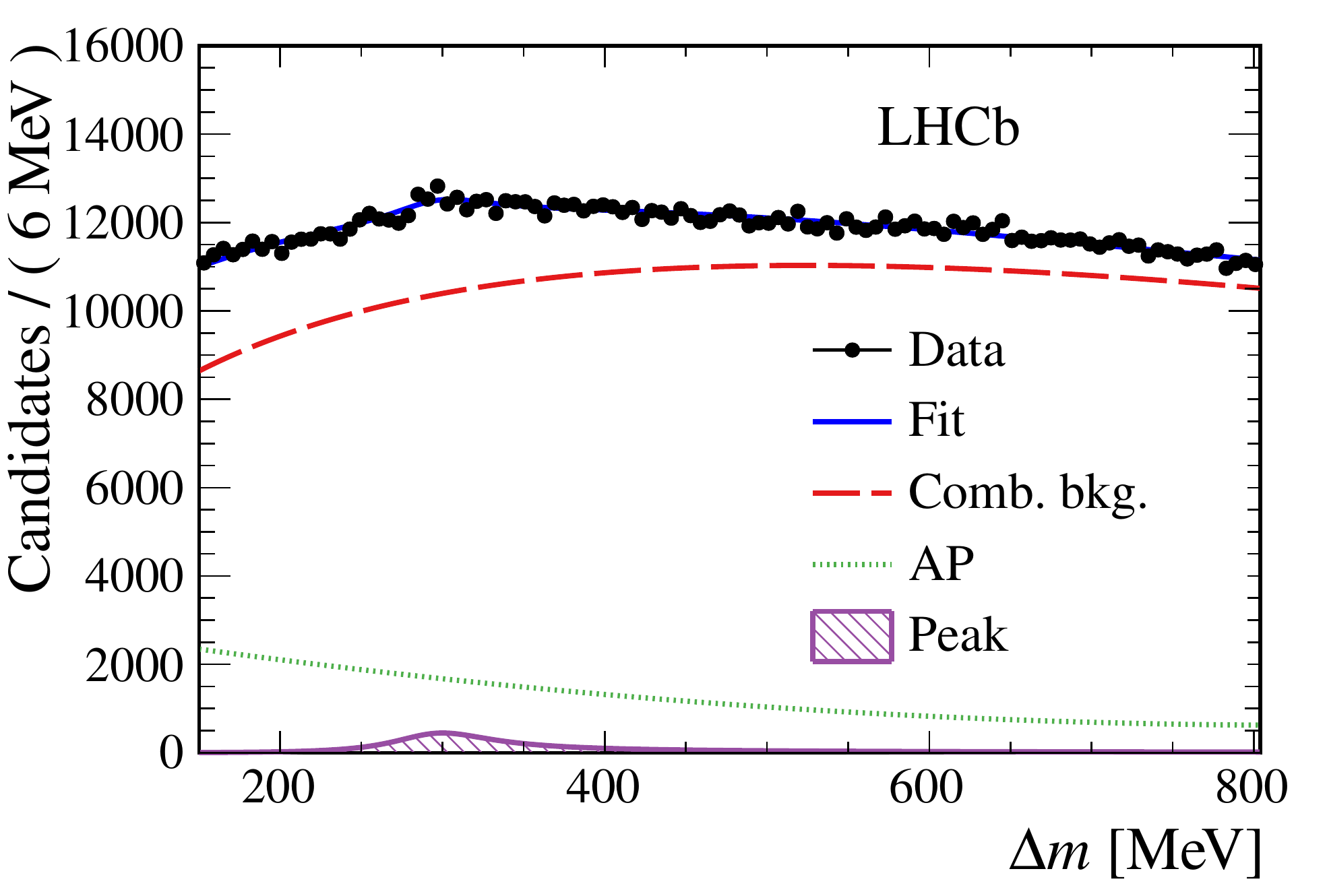}
	\end{subfigure}
	\begin{subfigure}{0.5\textwidth}
		\includegraphics[width=\textwidth]{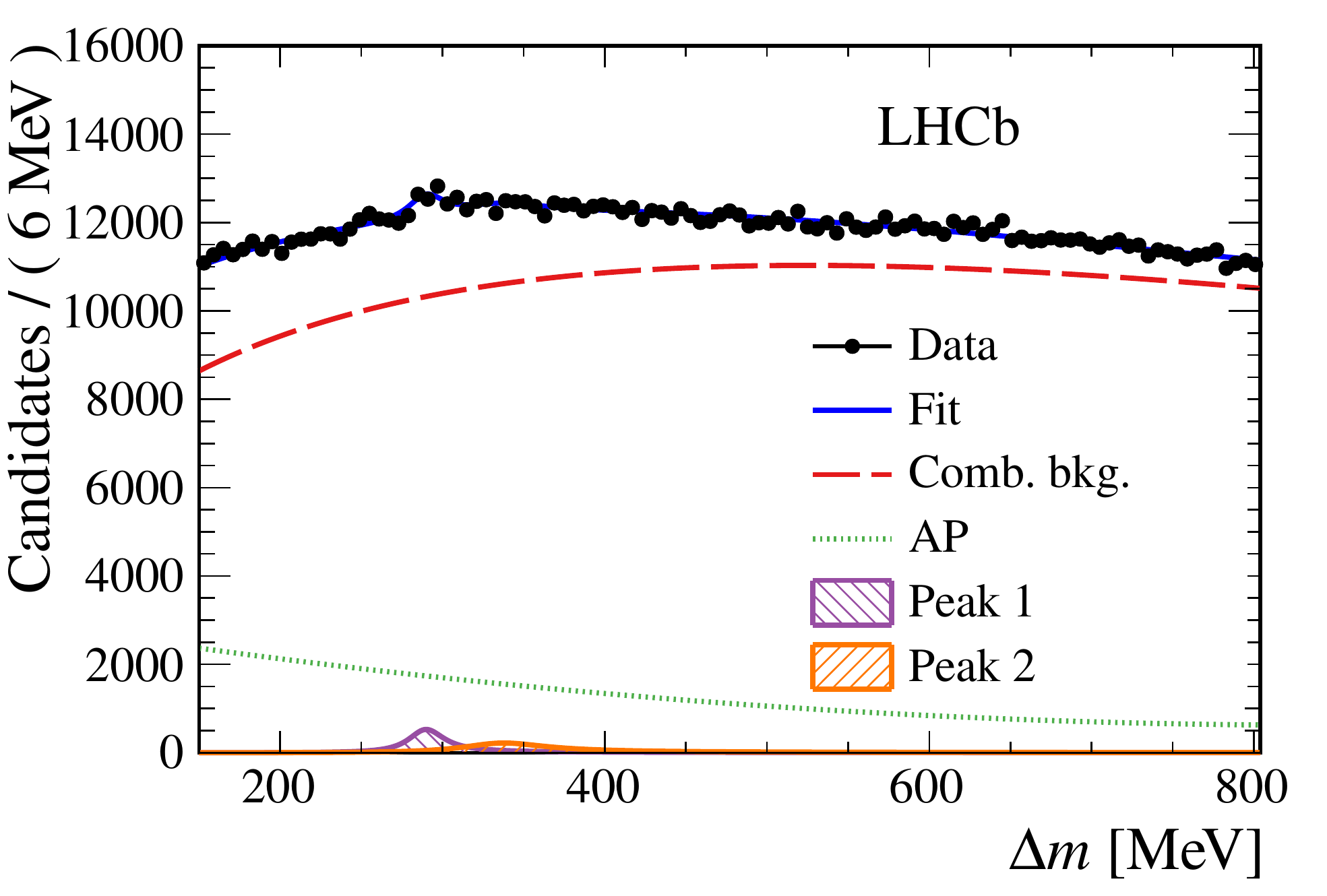}
	\end{subfigure} \\
	\begin{subfigure}{0.5\textwidth}
		\includegraphics[width=\textwidth]{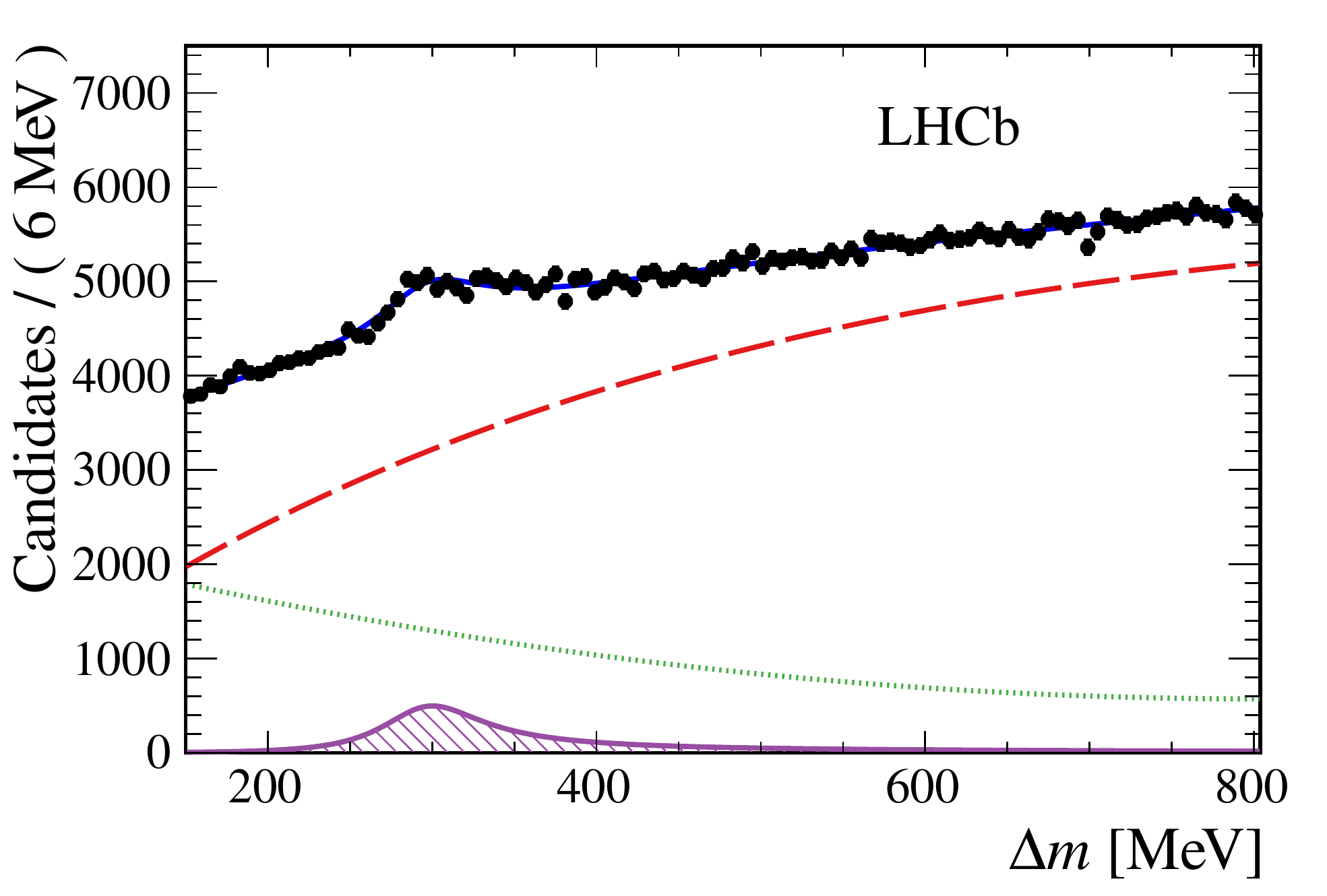}
	\end{subfigure} 
	\begin{subfigure}{0.5\textwidth}
		\includegraphics[width=\textwidth]{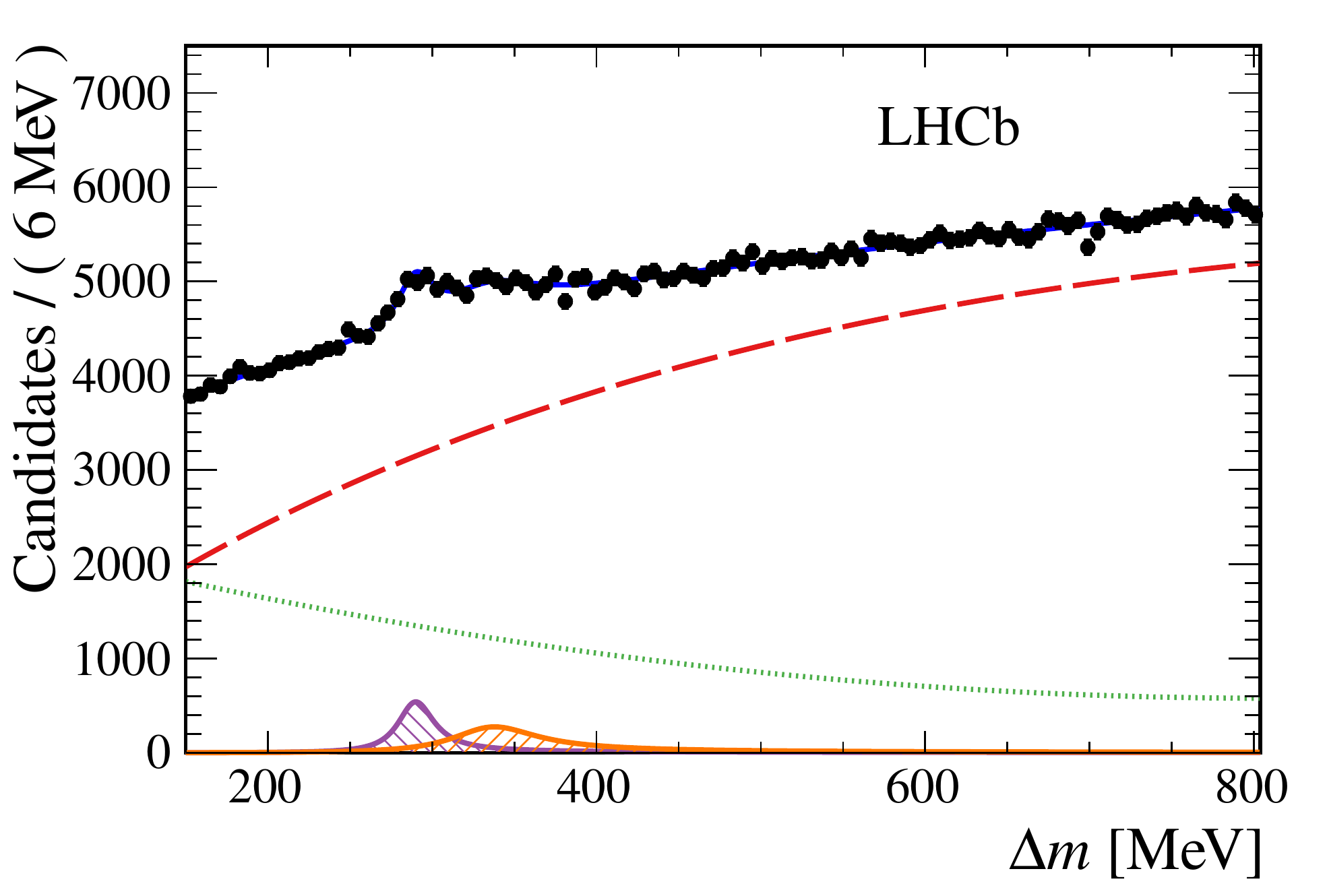}
	\end{subfigure} \\
	\begin{subfigure}{0.5\textwidth}
		\includegraphics[width=\textwidth]{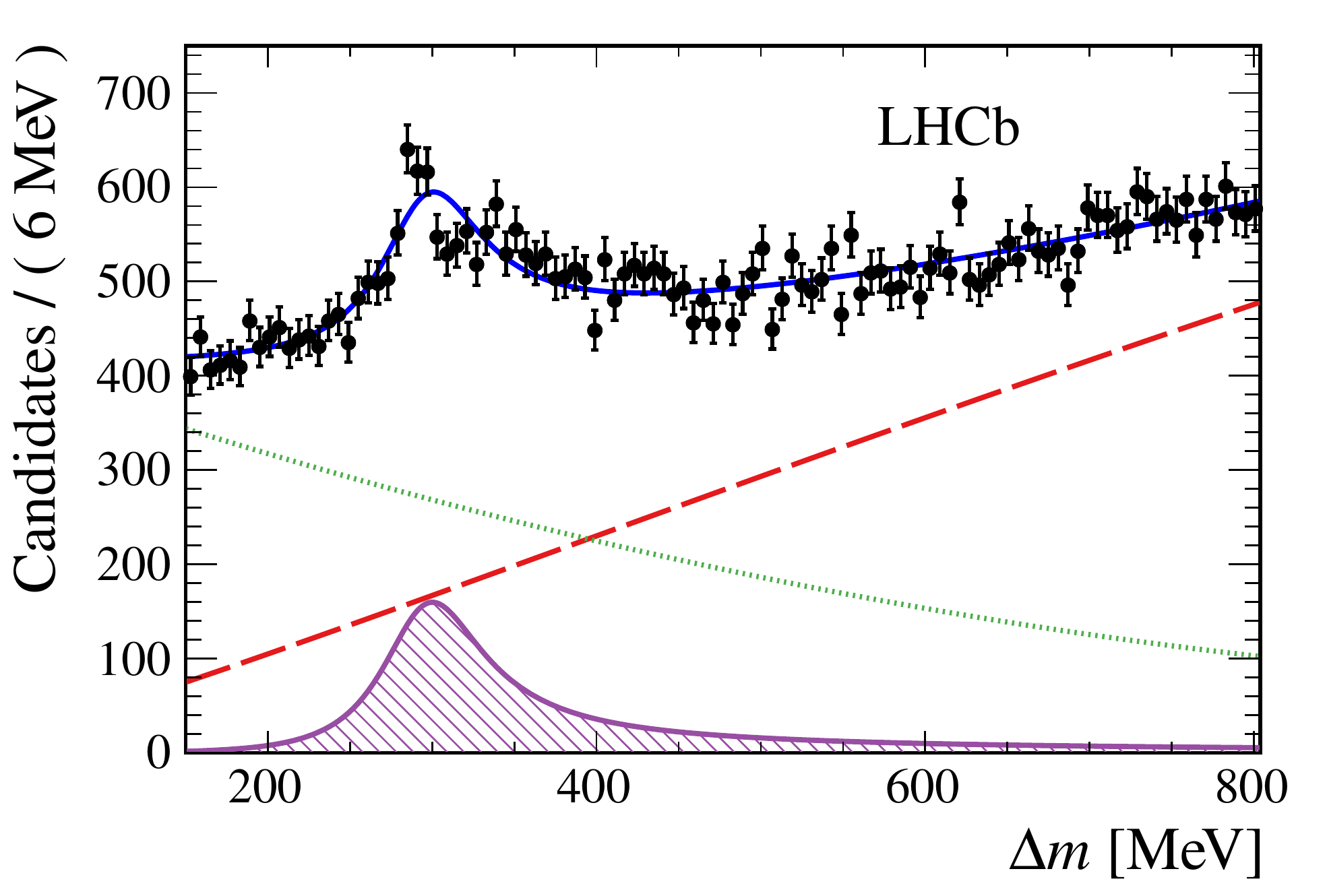}
	\end{subfigure}
	\begin{subfigure}{0.5\textwidth}
		\includegraphics[width=\textwidth]{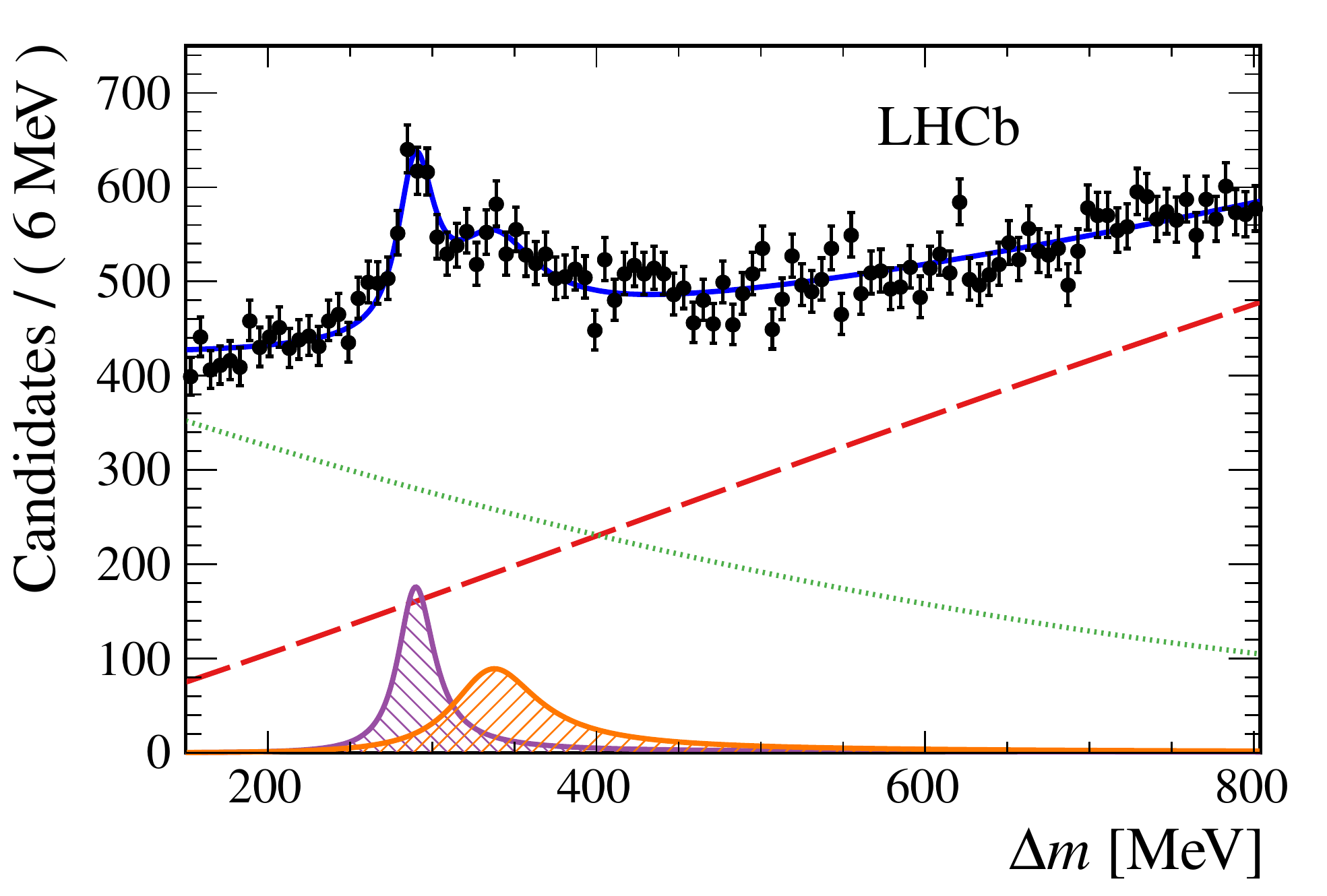}
	\end{subfigure}
\caption{The \Bu\Km mass difference distributions in data, overlaid with the fit models: (left column) one-peak hypothesis and (right column) two-peak hypothesis. In each column, the rows are from top to bottom for candidates with the prompt kaon \pt: $0.5 < \pt < \SI{1}{\giga\evolt}$, $1 < \pt < \SI{2}{\giga\evolt}$, and $\pt > \SI{2}{\giga\evolt}$. The legend in the top row applies for each column. The associated production (AP) background is described by a second-degree polynomial in each fit. The combinatorial background shape is fixed from a fit to the \Bu\Kp mass difference distributions.
  \label{fig:fit}}
\end{figure}

We use two resolution models: one where the decay goes directly to a \Bu\Km final state and the other where it proceeds through an intermediate \bstp meson which decays to $\Bu\gamma$. We model the resolution in the direct \Bu\Km decay as the sum of a Gaussian distribution and a Crystal Ball function~\cite{Skwarnicki:1986xj} with a shared mean, and in the $B^{*+}\Km$ channel as the sum of two Gaussian distributions with different means to account for the smearing and shift in the peak position by approximately \SI{45}{\mega\evolt} due to the unreconstructed photon. The detector resolution is small compared to the width of the prospective states; the effect of the missing photon is significant. The peak positions and widths are shared in each bin of the kaon \pt, but the yield is allowed to vary independently. The fit results using the polynomial AP shape are shown in \cref{fig:fit}. A total signal yield of \num{18900 \pm 2200} is observed in the two peaks across all the \pt bins when fit with the default model and \Bu\Km resolution.

We determine the local statistical significance of the observation using the asymptotic formula for the profile likelihood ratio test statistic~\cite{Wilks:1938dza}. We compare first the one-peak hypothesis to the null hypothesis, and then the two-peak to the one-peak hypothesis. The null fit model corresponds to only a polynomial description of the AP in addition to the shape of the same-sign data. The minimum of the test statistics across each fit with different AP descriptions (discussed later) gives local significances of more than $20\sigma$ for the one-peak fit with respect to the null hypothesis and $7.7\sigma$ for the two-peak description with respect to the one-peak hypothesis. As the significance is large, we have not made an exact quantification of the penalty due to the look-elsewhere effect.

Assuming the two-peak hypothesis, the masses and widths of the two states and the fraction of the yield in the lower mass state, $f_1$, are determined to be
\begin{align*}
    m_1      &= 6063.5 \pm   1.2 \text{ (stat)} \pm   0.8\text{ (syst)}\mev, \\
\Gamma_1 &=  26 \pm   4 \text{ (stat)} \pm   4\text{ (syst)}\mev, \\
m_2      &= 6114 \pm   3 \text{ (stat)} \pm   5\text{ (syst)}\mev, \\
\Gamma_2 &=  66 \pm   18 \text{ (stat)} \pm   21\text{ (syst)}\mev, \\
f_1 &= 0.47 \pm 0.11 \text{ (stat)} \pm 0.15\text{ (syst)},
\end{align*}
if the decay is directly to the \Bu\Km final state. If it instead proceeds through $B^{*+}\Km$, they are
\begin{align*}
m_1      &= 6108.8 \pm   1.1\text{ (stat)} \pm   0.7\text{ (syst)}\mev, \\
\Gamma_1 &=  22 \pm   5\text{ (stat)} \pm   4\text{ (syst)}\mev, \\
m_2      &= 6158 \pm   4\text{ (stat)} \pm   5\text{ (syst)}\mev, \\
\Gamma_2 &=  72 \pm   18\text{ (stat)} \pm   25\text{ (syst)}\mev, \\
f_1 &= 0.42 \pm 0.11\text{ (stat)} \pm 0.16\text{ (syst)}.
\end{align*}
 
The dominant contribution to the systematic uncertainty arises from the considered variations in the fit functions, and is obtained by repeating the fits with different conditions and calculating the differences with the default fit. We refit the data sample using alternative functional forms for the AP background: an exponential and a separate threshold function. Each AP description produces consistent results for the significance, with variation in the peak parameters. The dependence on the signal model is estimated by alternatively using relativistic Breit-Wigner models with different angular-momentum hypotheses of $\ell=2$ or $\ell=1$ and taking the largest absolute difference. We also vary the \pt binning and mass range for the fit, and we include the effect on one peak's parameters of changing the other peak from the \Bu\Km model to the $B^{*+}\Km$ model or vice versa. A breakdown of the systematic uncertainties is given in \cref{tab:param_systs}.

\begin{table}[tb]
\caption{Sources of systematic uncertainty on the peak parameters assuming a two-peak structure decaying either directly to $\Bu\Km$ or through $B^{*+}\Km$. The parameter differences with respect to the nominal fit result are given in \mev.\label{tab:param_systs}}
\begin{center}
    \begin{tabular}{ccSSSSS}
    \toprule
    Decay hypothesis & Source & \multicolumn{1}{c}{$m_1$} & \multicolumn{1}{c}{$\Gamma_1$} & \multicolumn{1}{c}{$m_2$} & \multicolumn{1}{c}{$\Gamma_2$} & \multicolumn{1}{c}{$f_1$} \\
    \midrule
    \multirow{3}{*}{$\Bu\Km$} & Background model & 0.4 & 1.2 & 0.7 & 10.6 & 0.07 \\
    & Signal model & 0.5 & 3.9 & 4.3 & 11.5 & 0.11 \\
    & Mass and \pt range & 0.4 & 1.3 & 1.6 & 13.6 & 0.07 \\
    \midrule
    & Total & 0.8 & 4.2 & 4.6 & 20.7 & 0.15 \\
    \midrule
    \multirow{3}{*}{$B^{*+}\Km$} & Background model & 0.4 & 1.7 & 0.7 & 13.1 & 0.08 \\
    & Signal model & 0.2 & 3.3 & 4.2 & 9.0 & 0.08 \\
    & Mass and \pt range & 0.5 & 2.1 & 1.3 & 19.5 & 0.10 \\
    \midrule
    & Total & 0.7 & 4.3 & 4.5 & 25.1 & 0.16 \\
    \bottomrule
    \end{tabular}
\end{center}
\end{table}

Since the mass difference between the two peaks is found to be close to the $B^{*+}$--\Bu mass difference of approximately \SI{45}{\mega\evolt}, we consider also the possibility that the structure is produced by a single resonance that decays in both the \Bu\Km and $B^{*+}\Km$ channels. In an alternative fit, both resolution models are simultaneously combined with a resonance lineshape described by a single mass and width. This hypothesis is disfavored by more than $2\sigma$ with respect to the two-state hypothesis and cannot be completely ruled out. Qualitatively, this model does not describe the data as well because the observed width is much different in the two peaks, with the lower peak being narrower.

%% file: Production.tex
\section{Production ratio}

In addition to the masses and widths of the new mesons, we also determine the production ratio relative to the \bstwost meson. The \Bu\Km candidates in both mass ranges are selected as previously described, but we use only candidates selected from the \decay{\Bu}{\jpsi\Kp} decay and require  the candidate to have triggered the event. We define the production ratio as the product of the cross-sections times branching fractions of the new states divided by the corresponding product for \bstwost,
\begin{equation}
  R = \frac{ \sum \sigma\qty(\bsstst) \times \BF( \decay{\bsstst}{\B^{(*)+}\Km} ) }{ \sigma\qty(\bstwost)\times \BF(\decay{\bstwost}{\Bu\Km} ) }.
\end{equation}
We sum the measured yield for both observed states, and consider the yield difference of the $\Bu\Km$ and $\bstp\Km$ models as a systematic uncertainty. The ratio is taken with respect to the yield for only the \Bu\Km decay of the \bstwost meson. The production cross-section is restricted to the fiducial region of \bsstst transverse momentum and rapidity, $0 < \pt < \SI{20}{\giga\evolt}$ and $2.1 < y < 4.2$.

The relative efficiency for a state at the observed mass with respect to the \bstwost meson is determined using simulation, and used to correct the observed yields. The largest systematic uncertainty on the relative efficiency comes from the unknown transverse momentum spectra of the new states. Additional efficiency uncertainties, for example from the description of particle identification in simulation, give smaller contributions. The \bstwost yield is determined from a separate unbinned maximum-likelihood fit in the range $50 < \Delta m < \SI{85}{\mega\evolt}$, using the same prompt kaon \pt binning as in the main fit. The systematic uncertainties from signal and background fit-model variations are handled as in the fit used for the observation. The fit is additionally separated into different data-taking periods since the efficiency and production ratio may depend on the collision energy and detector conditions.  The results, however, are statistically compatible between the different periods. We therefore quote the combined production ratio 
  \begin{equation*}
    R = 0.87 \pm 0.15 \text{ (stat)} \pm 0.19 \text{ (syst)}.
  \end{equation*}
A breakdown of the systematic uncertainties is given in \cref{tab:prod_systs}.

\begin{table}[tb]
\caption{Summary of the systematic uncertainties on the production ratio. The effect of the unknown \pt distribution is separated from other uncertainties affecting the relative efficiency.\label{tab:prod_systs}}
\begin{center}
\begin{tabular}{cc}
\toprule
Source & Absolute uncertainty \\
\midrule
Signal fit variations & 0.10 \\
Background fit variations & 0.10 \\
Mass and \pt range variations & 0.13 \\
\bsstst \pt distribution & 0.02 \\
Other efficiency uncertainties & 0.02 \\
\midrule
Total & 0.19 \\
\bottomrule
\end{tabular}
\end{center}
\end{table}

%% file: Conclusion.tex
\section{Conclusions}

An excess is observed in the \Bu\Km mass spectrum at a mass approximately 300\mev above the \Bu\Km threshold, which we interpret as resulting from the decays of multiple excited \bsstst mesons. The structure is not well described by a single resonance; the significance for a two-peak structure relative to a one-peak hypothesis is greater than $7\sigma$. A single resonance which can decay through both the \Bu\Km and \bstp\Km channels is disfavoured but cannot be excluded. Future investigation with a larger data sample and with the  \bstp\Km final state fully reconstructed could potentially better determine the exact structure in this mass range.
Under the two-peak hypothesis, we have measured the masses and widths of the two states. Based on predictions for the masses of excited \bsst states, the new observation is likely to result from $L=2$ orbitally excited mesons.  We have additionally determined the production cross-section times branching fraction of the new excess relative to the previously observed $L=1$ \bstwost, which shows that it is produced at a comparable rate.

%% file: acknowledgements.tex
\section*{Acknowledgements}
%
% These Acknowledgements valid from 3-May-2019
%
\noindent We express our gratitude to our colleagues in the CERN
accelerator departments for the excellent performance of the LHC. We
thank the technical and administrative staff at the LHCb
institutes.
We acknowledge support from CERN and from the national agencies:
CAPES, CNPq, FAPERJ and FINEP (Brazil); 
MOST and NSFC (China); 
CNRS/IN2P3 (France); 
BMBF, DFG and MPG (Germany); 
INFN (Italy); 
NWO (Netherlands); 
MNiSW and NCN (Poland); 
MEN/IFA (Romania); 
MSHE (Russia); 
MICINN (Spain); 
SNSF and SER (Switzerland); 
NASU (Ukraine); 
STFC (United Kingdom); 
DOE NP and NSF (USA).
We acknowledge the computing resources that are provided by CERN, IN2P3
(France), KIT and DESY (Germany), INFN (Italy), SURF (Netherlands),
PIC (Spain), GridPP (United Kingdom), RRCKI and Yandex
LLC (Russia), CSCS (Switzerland), IFIN-HH (Romania), CBPF (Brazil),
PL-GRID (Poland) and OSC (USA).
We are indebted to the communities behind the multiple open-source
software packages on which we depend.
Individual groups or members have received support from
AvH Foundation (Germany);
EPLANET, Marie Sk\l{}odowska-Curie Actions and ERC (European Union);
A*MIDEX, ANR, Labex P2IO and OCEVU, and R\'{e}gion Auvergne-Rh\^{o}ne-Alpes (France);
Key Research Program of Frontier Sciences of CAS, CAS PIFI,
Thousand Talents Program, and Sci. \& Tech. Program of Guangzhou (China);
RFBR, RSF and Yandex LLC (Russia);
GVA, XuntaGal and GENCAT (Spain);
the Royal Society
and the Leverhulme Trust (United Kingdom).

%% file: LHCb_Authorship_23-Jul-2020.tex
% LHCb collaboration author list
% Data extracted on October 12th, 2020 at 5:39pm for reference date 23-Jul-2020
\centerline
{\large\bf LHCb collaboration}
\begin
{flushleft}
\small
R.~Aaij$^{31}$,
C.~Abell{\'a}n~Beteta$^{49}$,
T.~Ackernley$^{59}$,
B.~Adeva$^{45}$,
M.~Adinolfi$^{53}$,
H.~Afsharnia$^{9}$,
C.A.~Aidala$^{84}$,
S.~Aiola$^{25}$,
Z.~Ajaltouni$^{9}$,
S.~Akar$^{64}$
% ...

{\footnotesize \it

% ...
$ ^{9}$Universit{\'e} Clermont Auvergne, CNRS/IN2P3, LPC, Clermont-Ferrand, France\\
% ...
$ ^{25}$INFN Sezione di Milano, Milano, Italy\\
% ...
$ ^{31}$Nikhef National Institute for Subatomic Physics, Amsterdam, Netherlands\\
% ...
$ ^{45}$Instituto Galego de F{\'\i}sica de Altas Enerx{\'\i}as (IGFAE), Universidade de Santiago de Compostela, Santiago de Compostela, Spain\\
% ...
$ ^{49}$Physik-Institut, Universit{\"a}t Z{\"u}rich, Z{\"u}rich, Switzerland\\
% ...
$ ^{53}$H.H. Wills Physics Laboratory, University of Bristol, Bristol, United Kingdom\\
% ...
$ ^{59}$Oliver Lodge Laboratory, University of Liverpool, Liverpool, United Kingdom\\
% ...
$ ^{64}$University of Cincinnati, Cincinnati, OH, United States\\
% ...
$ ^{67}$Syracuse University, Syracuse, NY, United States\\
% ...
$ ^{84}$University of Michigan, Ann Arbor, United States, associated to $^{67}$\\
\bigskip
% ...
\medskip
}
\end{flushleft}